\documentstyle[12pt,epsf]{article}

\newcommand{\itep}{~\vspace{-1.5cm}\begin{flushright}
{\large ITEP-TH-1/97}\end{flushright}\vspace{-.5cm}}
\newcommand{\talk}{\setcounter{footnote}{0}
\renewcommand{\thefootnote}{\fnsymbol{footnote}}
\footnote{Talk given by M.I.~Polikarpov
at 31st International Symposium Ahrenshoop
on the Theory of Elementary Particles
Buckow, September 2-6, 1997.}}

\def\1ad{\mbox{\normalsize $^1$}}
\def\2ad{\mbox{\normalsize $^2$}}
\def\3ad{\mbox{\normalsize $^3$}}
\def\4ad{\mbox{\normalsize $^4$}}
\def\5ad{\mbox{\normalsize $^5$}}
\def\6ad{\mbox{\normalsize $^6$}}
\def\7ad{\mbox{\normalsize $^7$}}
\def\8ad{\mbox{\normalsize $^8$}}
\def\makefront{\vspace*{1cm}\begin{center}
\def\newtitleline{\\ \vskip 5pt}
{\Large\bf\titleline}\\
\vskip 1truecm
{\large\bf\authors}\\
\vskip 5truemm
\addresses
\end{center}
\vskip 1truecm
{\bf Abstract:}
\abstracttext
\vskip 1truecm}
\newcommand{\eq}[1]{(\ref{#1})}
\newcommand{\diff}{\partial}

\newcommand{\beq}{\begin{equation}}
\newcommand{\eeq}{\end{equation}}
\newcommand{\beqn}{\begin{eqnarray}}
\newcommand{\eeqn}{\end{eqnarray}}

\def\NP{{\it Nucl.~Phys.}}
\def\PL{{\it Phys.~Lett.}}
\def\PRL{{\it Phys.~Rev.~Lett.}}
\def\PRp{{\it Phys.~Rep.}}
\def\PR{{\it Phys.~Rev.}}

\begin{document}
\itep
\def\titleline{Properties of Abelian Monopoles
\newtitleline
in $SU(2)$ Lattice Gluodynamics\talk
}
\def\authors{B.L.G.~Bakker \1ad, M.N.~Chernodub \2ad,
F.V.Gubarev \2ad,
M.I.~Polikarpov \2ad and A.I.~Veselov \2ad}
\def\addresses{
\1ad
Department of Physics and Astronomy, Vrije Universiteit,\\
De Boelelaan 1081, NL-1081 HV Amsterdam, The Netherlands\\
\2ad
ITEP, B.Cheremushkinskaya 25, Moscow, 117259, Russia
}
\def\abstracttext{
We discuss some properties of  abelian monopoles in
the Maximal Abelian projection of the  $SU(2)$ lattice  gluodynamics.
We show that in the maximal abelian projection
abelian monopoles carry fluctuating electric charge and that
the monopole currents are correlated with the magnetic and the
 electric parts
of the  $SU(2)$ action density.}

\makefront
\section{Introduction}
Abelian monopoles play a key role in the dual superconductor
mechanism of confinement~\cite{MatH76} in non-abelian gauge theories.
Abelian monopoles appear after the
so called abelian projection~\cite{tHo81}. Condensation of abelian
monopoles gives rise to the formation of an  electric flux tube between
the test quark and antiquark. Due to a non-zero string tension
the quark and the  antiquark are confined by a linear potential.
There are many  numerical facts~\cite{LatRew} which show that the abelian
monopoles in the Maximal Abelian (MaA) projection are responsible for the
confinement. The monopole condensation in the confinement phase of
gluodynamics has been established
 by the  investigation of various monopole creation
operators~\cite{MonopoleCond} in the MaA
projection~\cite{KrScWi87}. The $SU(2)$ string tension is
well described by the contribution of the abelian monopole
currents~\cite{MonopoleDominance}; these currents satisfy the London
equation for a superconductor~\cite{SiBrHa93}.

Below we discuss several recently found properties of the abelian monopole
currents. In Section~2 we show that in the vacuum of the $SU(2)$ lattice
gluodynamics the abelian monopoles currents are correlated with the electric
currents. In Section~3 we show that the abelian monopoles are locally
correlated with electric and magnetic parts of the
 $SU(2)$ action density. All
numerical calculations are performed in the MaA projection.

\section{Abelian Monopoles Carry Electric Charge}

Consider a (anti-) self--dual configuration of the $SU(2)$ gauge field:
\beqn
F_{\mu\nu}(A)=\pm \frac{1}{2} \varepsilon_{\mu\nu\alpha\beta}
F_{\alpha\beta}(A) \equiv \pm \tilde{F}_{\mu\nu}\,,
\label{duality}
\eeqn
where $F_{\mu\nu}(A) = \partial_{[\mu,} A_{\nu]} + i [A_\mu,A_\nu]$.
In the MaA projection the commutator term \mbox{${\rm Tr} (\sigma^3
[A_{\mu},A_{\nu}])$} of the field strength tensor $F^3_{\mu\nu}$ is
suppressed, since the MaA projection is defined~\cite{KrScWi87} by
the minimization of the functional $R[A] = \int d^4 x
[(A_{\mu}^{1})^2 + (A_{\mu}^{2})^2]$ over the gauge transformations.
Thus,  in the said  projection, the fields $A_\mu (x)$ are as close
to abelian (diagonal) fields  as possible. Therefore,  in the MaA
projection eq.(\ref{duality}) yields~\cite{BornSchierholz}:
$f_{\mu\nu}(A)=\partial_\mu A^3_\nu - \partial_\nu A^3_\nu \approx\pm
{\tilde f}_{\mu\nu}(A)$.  Thus, the abelian monopole currents must be
accompanied by the electric currents:  $J^e_{\mu}=\diff_{\nu}
f_{\mu\nu}(A)\approx \pm\diff_{\nu}{\tilde f}_{\mu\nu}(A)=\pm
J^m_{\mu}$.  Therefore, in the MaA projection the abelian monopoles
are dyons for (anti) self-dual $SU(2)$ field
configurations~\cite{BornSchierholz}. Below we show that in the real
(not cooled) vacuum of lattice gluodynamics the abelian monopole
currents are correlated with the electric currents~\cite{jejm}.

In order to study the relation of electric and magnetic currents, we
have to calculate  connected  correlators of these currents.  The
simplest correlator $\ll J^m_{\mu}J^e_{\mu} \gg \equiv $
$<J^m_{\mu}J^e_{\mu}> - < J^m_{\mu}> <J^e_{\mu}>$ is zero, since
$<J^m_{\mu}J^e_{\mu}> = 0$ due to the opposite parities of the
operators $J^m$ and $J^e$,  and   $<J^{m,e}_{\mu}>=0$ due to the
Lorentz invariance. The simplest non--trivial (normalized) correlator
is
\beqn
{\bar G } =
\frac{1}{\rho^e \rho^m} \, <J^m_{\mu}(y) J^e_{\mu}(y) q(y)>\,,
\label{G}
\eeqn
where $q(x)$ is the sign of the topological charge density at the
point $x$ and $\rho_{m,e} = $ $ \sum_{l} <|J^{m,e}_l|> \slash (4V)$
are the densities of the magnetic and the electric charges, $V$ is
the lattice volume (total number of sites).

We perform a numerical calculation of the  correlator \eq{G} in the
$SU(2)$ lattice gauge theory on the $8^4$ lattice with periodic
boundary conditions. We use  100 statistically independent gauge
field configurations for each value of $\beta$.

\begin{figure}[htb]
\begin{center}
\begin{tabular}{cc}
{\epsfxsize=0.47\textwidth\epsfbox{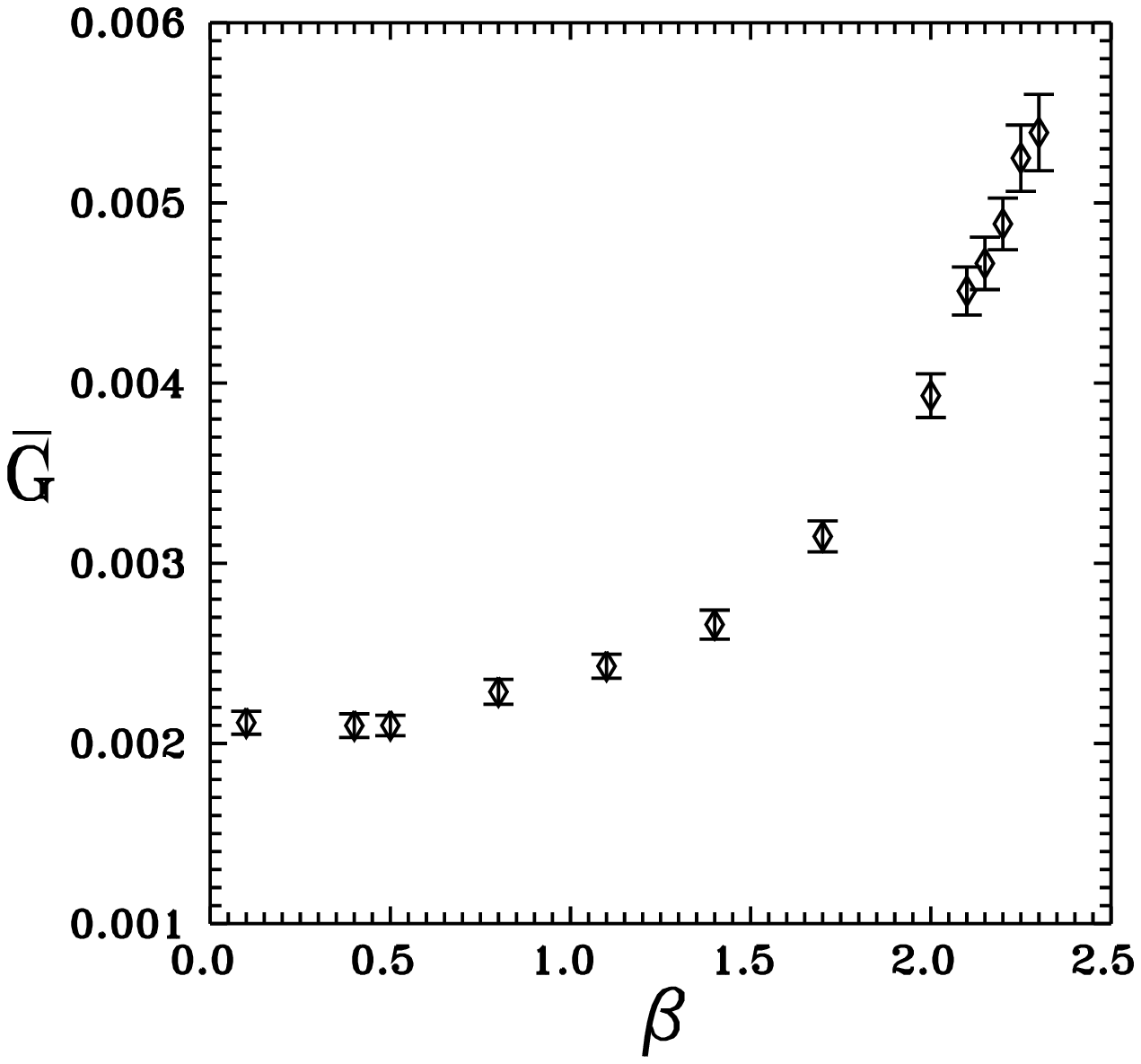}}&
{\epsfxsize=0.45\textwidth\epsfbox{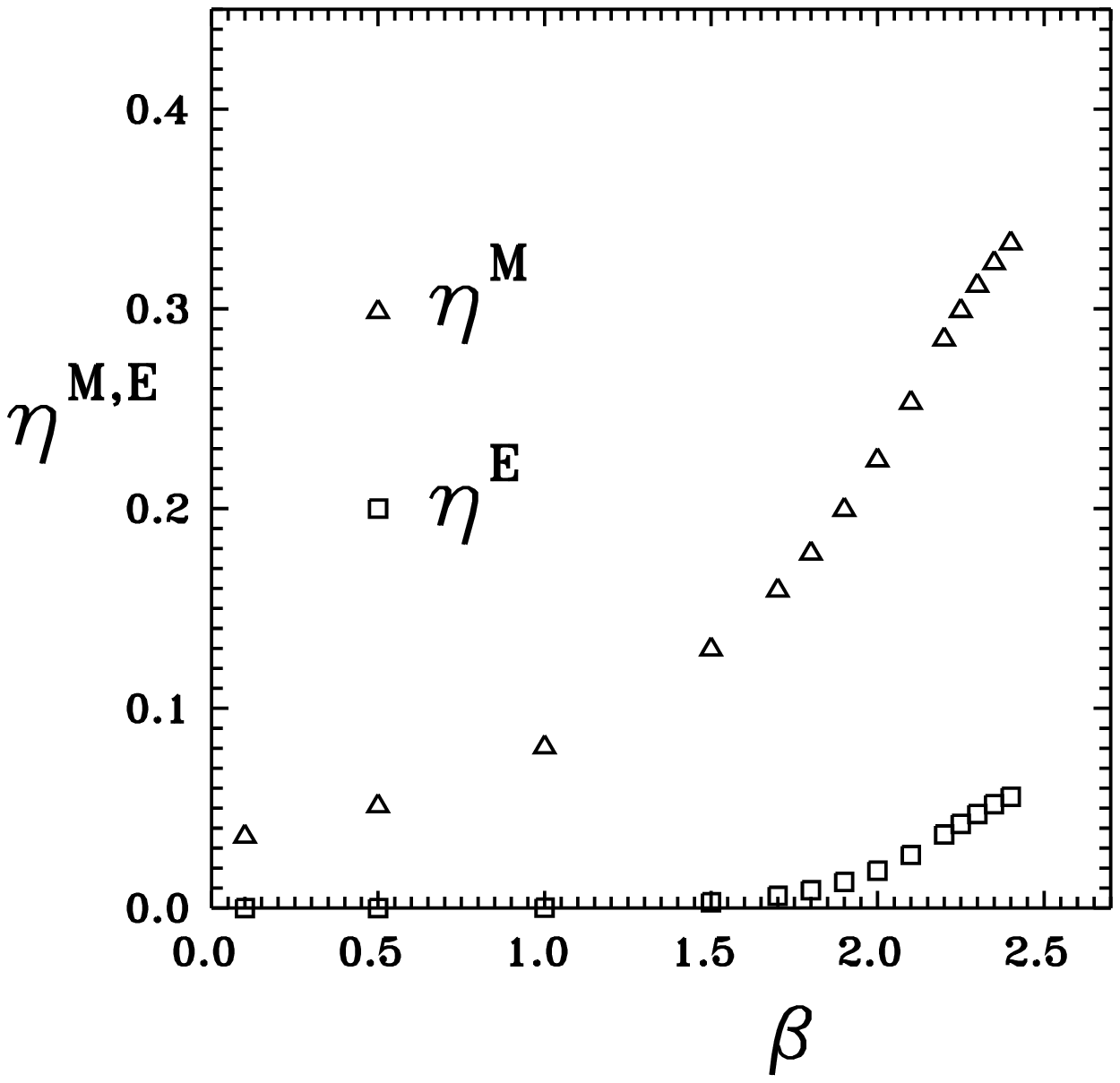}}\\
 (a) & (b)\\
\end{tabular}
\end{center}
\caption{(a) The dependence of the correlator ${\bar G}$ on $\beta$;
~(b) The relative excess of the magnetic (circles, from
Refs.~\cite{BaChPo97}) and the electric (boxes) action density near
the monopole current.  The data are extrapolated to the infinite
lattice size.}
\label{one}
\end{figure}

The dependence of the correlator ${\bar G}$ on $\beta$ is shown in
Fig.~\ref{one}(a). This correlator is positive for  all values of
$\beta$. Therefore, the abelian monopoles in the MaA
projection carry  an electric charge, too. According to definition \eq{G},
the sign of the electric charge of the monopole coincides with the product
of the magnetic charge and the topological charge. Thus, in the
gluodynamic vacuum the abelian monopoles become abelian dyons due to a
non-trivial topological structure of the vacuum gauge fields.

\section{Abelian Monopole Currents are Correlated with $SU(2)$ Action
Density}

Abelian monopoles appear as singularities in the gauge
transformations~\cite{tHo81,LatRew}. On the other hand, the monopole
currents reproduce the $SU(2)$ string
tension~\cite{MonopoleDominance}. Thus, monopoles  are likely to be
related to some physical objects. A physical object is something
which carries action. Below we study the local correlations of the
abelian monopoles with the density of the magnetic and the electric
parts of the $SU(2)$ action (the global correlation was found in
Ref.~\cite{ShSu95}). We show that the monopoles are physical objects
but it does not mean that these have to propagate in the Minkowski
space; a chain of instantons can produce a  similar effect: an
enhancement of the action density along a line in  Euclidean space.
The simplest quantities which can show this correlation are the
relative excess of the magnetic and the electric action densities
$\eta^{M,E} = (S^{M,E}_m - S) \slash S$ in the region near the
monopole current.  Here $S$ is the expectation value of the lattice
plaquette action, $S_P = <( 1 - \frac 12 Tr \, U_{P})>$. The
quantities $S^{M,E}_m$ are, respectively, the magnetic and the
electric parts of the $SU(2)$ action density, which are calculated on
plaquettes closest to the monopole current.

In the continuum notation, the quantities $S^{M,E}_m$ have the following
form:
\beqn
S^M_m   =   \frac{1}{2} <{\rm Tr}{(
n_\mu (x)\,{\tilde F}_{\mu\nu}(x))}^2>\,,\quad
S^E_m   =   \frac{1}{2} <{{\rm Tr} {(n_\mu(x)\,
F_{\mu\nu}(x)})}^2 >\,,
\eeqn
$n_\mu (x)$ is the unit vector in the direction of the current:
$n_\mu (x)=j_\mu (x)/| j_\mu (x) |$, if $j_\mu (x) \neq 0$, and
$n_\mu (x)= 0$ if $j_\mu (x) = 0$. It is easy to see that for a
static monopole ($j_0 \neq 0; \,\, j_i=0, i=1,2,3 $) $S^M_m$ (resp.,
$S^E_m$) corresponds to the chromomagnetic action density
${(B^a_i)}^2$ (resp., chromoelectric action density ${(E^a_i)}^2$) at
the monopole current.

We calculate the quantities $\eta^{M}$ and $\eta^{E}$ on  symmetric
lattices $L^4$ of different lattice size  $L=8,10,12,16,20,24,30$ with
periodic boundary conditions. In Fig.~\ref{one}(b) we show the quantities
$\eta^{M,E}$ extrapolated to the infinite lattice size, ($L\to\infty$) $vs.$ $\beta$. The
monopole currents are calculated in the MaA projection. In
Fig.~\ref{one}(b) the statistical errors are smaller than the size of
the symbols. It is clearly seen that the abelian monopoles are
correlated with both the magnetic and the electric parts of the $SU(2)$
action density. Note that the correlation of the monopole charge with
the magnetic action density is larger than the correlation with the
electric part of the $SU(2)$ action.

\section*{Conclusion and Acknowledgments}

Our results show that the abelian monopoles in the MaA
projection of the $SU(2)$ gluodynamics {\it i)} have a fluctuating electric
charge; {\it ii)} carry the $SU(2)$ action.

This work was supported by the grants
INTAS-RFBR-95-0681, RFBR-96-1596740 and RFBR-96-02-17230a.

\end{document}